\documentclass[aps,prl,preprint,superscriptaddress,amsmath,amssymb]{revtex4}
\usepackage{graphicx}
\usepackage{color}
\begin{document}
\title{Strong-field ionization and fragmentation of large, gas-phase clusters in the few-cycle domain}
\author{D. Mathur}
\email{atmol1@tifr.res.in}
\affiliation{Tata Institute of Fundamental Research, 1 Homi Bhabha Road, Mumbai 400 005, India}
\author{F. A. Rajgara} 
\affiliation{Tata Institute of Fundamental Research, 1 Homi Bhabha Road, Mumbai 400 005, India}
\author{A. R. Holkundkar} 
\affiliation{Laser and Neutron Physics Section, Bhabha Atomic Research Centre, Mumbai 400 085, India}
\author{N. K. Gupta}
\affiliation{Laser and Neutron Physics Section, Bhabha Atomic Research Centre, Mumbai 400 085, India}

\begin{abstract}
Intense 3-cycle pulses (10 fs) of 800 nm laser light are utilized to measure energy distributions of ions emitted following Coulomb explosion of Ar$_n$ clusters ($n$=400-900) upon their irradiation by peak intensitis of 5$\times$10$^{14}$ W cm$^{-2}$. The 3-cycle pulses do not afford the cluster sufficient time to undergo Coulomb-driven expansion, resulting in overall dynamics that appear to be very different to those in the many-pulse regime. The peak ion energies are much lower than those obtained when 100 fs pulses of the same intensity are used; they are almost independent of the size of the cluster (over the range 400-900 atoms). Ion yields are a factor of 20 larger in the direction that is perpendicular to the laser polarization vector than along it. This unexpected anisotropy is qualitatively rationalized using molecular dynamics calculations in terms of shielding by an electronic charge cloud within the cluster that is spatially asymmetric.  
\end{abstract}
\maketitle

\section{Introduction}

Intense few-cycle optical pulses are of considerable significance in the contemporary pursuit of attosecond pulses and in studies of light-matter interactions in the domain of ultra-short and strong fields \cite{Brabec}. Different strategies for few-cycle pulse generation have been demonstrated, such as hollow-fiber pulse (HFP) compression \cite{Nisoli} and optical parametric amplification \cite{Kobayashi}. At present, HFP methods are the most widely adopted even though they present experimental difficulties vis-a-vis alignment and other practical limitations like low coupling efficiency. An alternate method based on filamentation in rare gases has also found recent utility \cite{filamentation} and pulses as short as 5 fs have been generated using two-stage filamentation in argon gas. This is the strategy that has been adopted in our laboratory \cite{our_few_cycles} to generate four-cycle pulses of intense 800 nm light which have been used by us to probe the ionization and fragmentation dynamics of molecules \cite{our_molecules} and to demonstrate a unimolecular bond rearrangment in H$_2$O on the timescale of a single vibrational period \cite{our_rearrangement}. We now report application of such few-cycle pulses to studies of the strong-field dynamics of gas-phase clusters of Ar. All hitherto-existing experimental work in the area of laser-cluster interactions in the strong field regime has been carried out using infrared laser pulses of 30 fs duration or longer ($>$10 optical cycles). Our experiments have used intense pulses of 800 nm laser light in the few-cycle domain (9-35 fs pulse durations) wherein time-of-flight spectrometry has been utilized to measure energy distributions of ions that are ejected upon Coulomb explosion of clusters Ar$_{400}$ - Ar$_{900}$ that become highly charged following irradiation in the intensity regime up to 5$\times$10$^{14}$ W cm$^{-2}$. The ionic products of Coulomb explosion of Ar-clusters in the few-cycle domain are seen to have significantly lower energy than is the case upon irradiation with many-cycle pulses. Moreover, the dependence of ion yield on the polarization direction of the incident laser pulse is different in the few-cycle and many-cycle regimes: in the few-cycle experiments we have conducted, twenty times more ions are produced when the incident polarization is in a direction that is perpendicular to the ion detection direction, in stark contrast to the results we obtain in the many-cycle (100 fs) domain \cite{vinod_asymmetry}. We rationalize this unexpected result by means of molecular dynamics calculations. Our experimental results throw new light on laser-cluster interactions in the ultrashort strong-field domain and open opportunities for further work that is qualitatively new. 

The course of the last fifteen years has seen the physics governing the interaction of laser fields with large gas-phase clusters develop into a rich and vibrant subset of research on the behaviour of matter in strong optical fields, and several reviews have been published \cite{Saalmann,Krainov,Mathur} that provide a cogent overview of different insights that have been gained. A major physics question that remains open after a decade and a half's worth of sustained experimental and theoretical work relates to the mechanism by which energy transfers from the optical field to the cluster. It is primarily electrons in the cluster that are the direct absorbers of optical energy; the absorbed energy is collisionally redistributed within the cluster so as to produce energetic ions, electrons and, at higher densities, intense incoherent radiation. The level of ionization and the mean energy of the electrons and ions are significantly higher than those obtained in strong-field ionization of single atoms and molecules. The hot nanoplasma that is created within the irradiated cluster undergoes spatial expansion and breaks up in a few picoseconds, yielding ions with kinetic energies that can be as high as a few MeV \cite{Ditmire}, up to five orders of magnitude larger than the energies obtained upon Coulomb explosion of single multiply-charged molecules \cite{MathurPhysRep}. Our measurements on cluster disassembly leading to ion production were made in the hitherto-unexplored temporal regime that is very substantially shorter than typical cluster expansion times. 

According to prevailing wisdom, collisional absorption (inverse bremsstrahlung) is the most likely energy-transfer mechanism which involves the transferring of energy from electrons to other particles in a series of inelastic collisions within the cluster nanoplasma that is created upon initial field-induced inner-ionization. In cluster parlance, `inner ionization' refers to ejection of electrons from the individual Ar-atoms that constitute the cluster in our experiments. Such electrons are quasi-free: they are not bound to any specific Ar-atom but they are spatially constrained within the cluster by the Coulombic charge of the positive ions. It is these spatially constrained electrons that are subjected to heating by the oscillating optical field and the ensuing collisions with plasma constituents can be thought of as being responsible for the ``bulk temperature" of the plasma. As further inner ionization occurs, the plasma expands due to Coulombic forces and its density becomes less. The build-up of ionic charge within the expanding cluster eventually leads to its explosion, whereupon energetic charged particles are emitted. Alternative mechanisms that might play a role in energy transfer are non-collisional in nature, such as resonance absorption \cite{Ginzburg} and vacuum (or Brunel) heating \cite{Brunel}. In our study we have attempted to disentangle the possible effect of these processes by taking recourse to few-cycle laser pulses with peak intensities that lie below 10$^{15}$ W cm$^{-2}$. This intensity regime allows us to discount resonance absorption while our temporal regime allows us to satisfactorily deal with a less-addressed facet of the collisional mechanism, namely the role played by enhanced ionization (EI) as a possible contributor to energy absorption by the cluster. The role of EI in clusters has not been explicitly addressed in any experimental study thus far, but Rost and coworkers \cite{Saalmann} have suggested that EI's influence on the dynamics could be probed by using a laser whose frequency was very much higher than the plasmon frequency of an unperturbed cluster wherein there would be no resonant absorption and energy would be transferred from the optical field into the cluster solely via EI. We adopt here the alternate strategy of `switching off' EI by the simple expedient of using laser pulses that are too short for significant nuclear motion to occur. Simultaneously, we focus attention on clusters that are large enough that EI might not be expected to be an important contributor to the overall dynamics. EI would be expected to contribute to the outer ionization of electrons when two or more ions within a cluster happen to align along the laser polarization vector. If there are shells of ions beyond the ones that thus line up, outer ionization is precluded. Consequently, EI is expected to be of decreasing importance as cluster size increases. 

We measured ion energies resulting from a range of Ar-clusters, whose mean size was varied from Ar$_{400}$ to Ar$_{900}$ by controlling the stagnation pressure behind the gas nozzle \cite{vinod_asymmetry}. Figure 1 shows typical results that we obtained when Ar$_{400}$, Ar$_{750}$, and Ar$_{900}$ clusters were irradiated with single pulses of 10 fs duration, with peak laser intensity of 5$\times$10$^{14}$ W cm$^{-2}$. We note three different facets of these energy distributions that distinguish them from those that are obtained when measurements were made with many-cycle laser pulses of the same intensity \cite{Saalmann}.

\begin{figure}
\includegraphics[width=10cm]{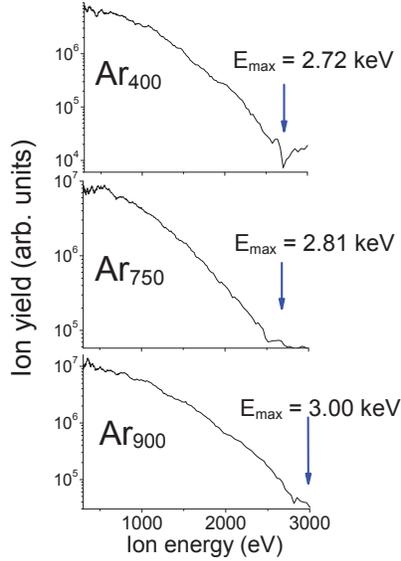}
\caption{(Color online) Energy distributions of ions obtained upon irradiation of Ar-clusters with 3-cycle (10 fs) pulses of 800 nm laser light of intensity 5$\times$10$^{14}$ W cm$^{-2}$. The laser polarization vector was aloigned perpendicular to the ion detection axis. Ion yields were a factor fo 20 smaller when the laser polarization vector was parallel to the ion detection axis.}
\end{figure}

Firstly, we note that maximum ion energies for these all clusters lie in the range 2.7-3.0 keV and it is clear that there is little variation with cluster size. The shapes of the energy distribution functions are noteworthy in that they differ significantly from shapes that are obtained in longer-pulse experiments. The second facet of our results relate to there being only a marginal dependence of the maximum ion energy on cluster size. This is in also in marked contrast to size-dependent ion energies that have been measured in the many-cycle (100 fs) temporal domain \cite{Mathur}. The third facet is possibly the most unexpected: our data for Ar-clusters shows a strong anisotropy of the ion energy spectrum with respect to the laser polarization vector such that the ion signal that we detected perpendicular to the laser polarization vector was much higher (typically 20-fold) than in the parallel direction. This is also in marked contrast to the findings in many-cycle experiments (see \cite{Saalmann} and references therein). 

On the first facet noted above, earlier work \cite{vinod_asymmetry,vinod2} has established that ion energy distributions obtained with 100 fs pulses comprise two distinct components. One component results in maximum values of ion energies that are consistent with what would be expected on the basis of Coulomb's law applied to a charged spherical cluster of a given size. This component is isotropic with respect to the incident polarization vector. Additionally, there is a second, higher energy ion component whose maximum energies extend well beyond the Coulomb limit. This high energy component exhibits anisotropy with respect to the laser polarization vector \cite{vinod_asymmetry,vinod2}. The two components are separated in measured energy distribution functions by a knee-like feature that is missing in the few-cycle data shown in Fig. 1. The angular anisotropy experimentally observed with 100 fs pulses was subsequently theoretically rationalized \cite{Saalmann} to be a consequence of polarization-induced effects that arise from the phase difference between the optical field and the oscillations of the negatively-charged cloud generated within the cluster by inner-ionized electrons. Such effects give rise to a ``flipping" of charges at the two poles of an initially spherical cluster \cite{vinod_asymmetry,vinod2}, with the poles being aligned along the laser's E-vector. However, such charge flipping might not be expected to be of consequence when cluster irradiation is by a 3-cycle pulse as there would be insufficient time for significant phase difference to be accumulated. The asymmetric ion and electron emission that has been reported upon cluster disassembly \cite{vinod_asymmetry,vinod2} would, therefore, not be expected when few-cycle pulses are employed. We discuss this more in teh following but, prior to that, we note the second facet of our results, namely the size independence of maximum ion energy that is observed in our three-cycle experiments. This is of interest from the viewpoint of results of calculations of Breizman and coworkers of electron response and ion acceleration in clusters that have been carried out using a nonlinear approach \cite{Breizman}. These calculations were carried out using two ways of describing the incident laser field and addressed cluster size issues in the following manner. For oscillating optical fields, which are more appropriate in the few-cycle case, it can be deduced from the treatment of Breizman {\it et al.} that only 2\% of electrons will actually leave Ar$_{900}$ and only 10\% will leave Ar$_{400}$. Consequently, one would expect relatively little change in the extent to which these clusters are collisionally heated as cluster size increases from 400 atoms to 900 atoms. Consequently, one would predict little dependence of maximum ion energies on cluster size. This, indeed, is the observation. On the other hand, if the optical field were to be treated as a static field, which would be more appropriate in the case of longer, many-cycle laser pulses, a much larger fraction of electrons would leave the cluster, varying between 53\% for Ar$_{400}$ to about 35\% for Ar$_{900}$, indicating that a larger variation in maximum ion energies might be expected as cluster size changes. Our results are, indeed, consistent with those that are theoretically expected when an oscillating optical field is made to act on cluster electrons in the theoretical treatment of Breizman {\it et al.} \cite{Breizman}.

We now address the third, and most significant, facet of our results, namely the very strong anisotropy of the ion energy spectrum with respect to the laser polarization vector. The ion yield that was measured was strongly anisotropic, with a 20-fold increase in yields when the ion signal that was detected perpendicular to the laser polarization vector. This is in stark contrast to earlier measurements of the angular distributions of energetic ions that are emitted upon explosion of Ar- and Xe-clusters irradiated with 100 fs long laser pulses \cite{vinod_asymmetry,vinod2}. We attempt to rationalize our observations in terms of differing shielding processes that occur in the few-cycle and many-cycle temporal regimes. We have taken recourse to model calculations of electron dynamics on ultrashort timescales using a molecular dynamics (MD) approach of the type described by Petrov and coworkers \cite{Petrov}. For computaitonal ease we consider here a single He$_{200}$ cluster of radius $R_0$ whose center is taken to lie at the origin of the coordinate system (x=y=z=0). The number of atoms, $N$, comprising the cluster size is $N=(R_0/R_W)^3$, where $R_W$ is the Wigner-Seitz radius. The presence of neighboring clusters is accounted for by imposing a periodic boundary conditions on the faces of our simulation cubical box with each side equal to the inter-cluster distance, $R_{IC}$ (in the present case, $R_{IC}=20~R_0$). Cluster ionization is implemented in Monte-Carlo fashion, using the oft-used ADK tunnel ionization formula \cite{ADK} applied to each ion and atom after each time step, $\Delta t$, for which we compute the force due to the optical field and the Coulomb force due to the presence of other charged particles. Each particle is then advanced to its new position according to its relativistic equations of motion. In Fig. 2 we focus on the spatial distribution of electrons produced within the cluster on a timescale of 10 fs. The Coulomb explosion occurs between 20 fs and 40 fs and prior to it, the build-up of the electronic charge cloud is seen to be asymmetric, with alignment along the direction of the laser polarization vector.

\begin{figure}
\includegraphics[width=6cm]{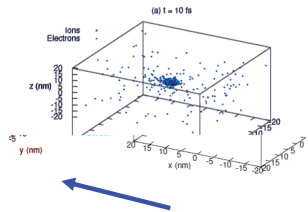}
\caption{(Color online) A snapshot of the asymmetric electron and ion spatial profiles in a 20 {\AA} helium cluster after irradiation by a 10 fs laser pulse. The arrow marks the direction of the laser polarization vector.}
\end{figure}

This asymmetric charge could results in shielding of ions within the core of the cluster such that the extent of shielding is spatially inhomogeneous: the polar regions of the speherical cluster (in the direction of the laser polarization vector) are much more effectively shielded than the equatorial regions. Consequently, when the ionic charge cloud in the core builds up sufficiently to result in a Coulomb explosion, the products of this explosion manifest themselves more prominently along the equatorial region than in the polar regions. This is preceisly what is observed in our experiments, with a 20-fold diminishing of ion yields in the laser polarization direction compared to in the orthogonal direction. The indications that emerge from the MD calculations that we take recourse to can only be regarded as rough guideposts to rationalization of the observed ionic anisotropy. It is clear that more elaborate calculations need to be undertaken in order to properly account for the shielding effects of the electronic charge cloud within the cluster. In long-pulse experiments, such shielding is probably of little consequence as the charge-flipping process referred to above dominates the ion dynamics and leads to the opposite type of aysmmetric emission of Coulomb explosin products, more along the polarization direction and perpendicular to it.

\begin{figure}
\includegraphics[width=10cm]{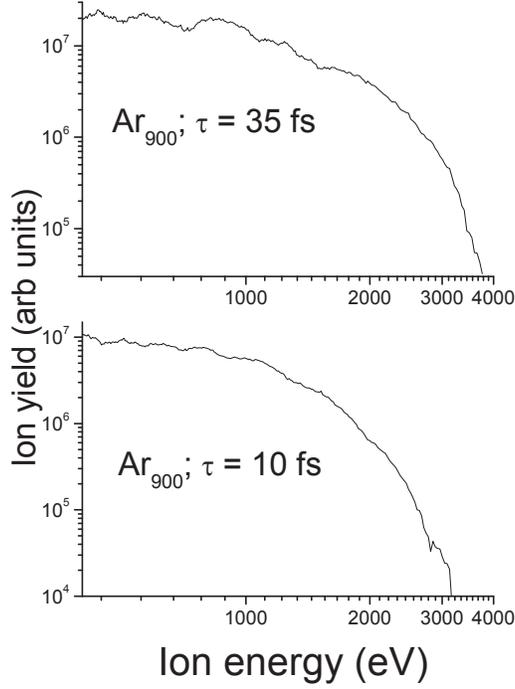}
\caption{(Color online) Ion energy spectra measured using pulses of 10 fs (top panel) and 35 fs (lower panel) duration, both of peak intensity 5$\times$10$^{14}$ W cm$^{-2}$.}
\end{figure}

We now consider in more detail the dependence of the ion energy distribution functions on pulse duration. Figure 3 shows energy spectra measured for pulses of 10 fs and 35 fs duration. The pulse energy was appropriately compensated in these two measuements to ensure that the peak intensity experienced by Ar$_{900}$ clusters was the same. The maximum energy obtained with 35 fs pulses is 3.9 keV, compared to 3.15 keV in the case of 10 fs pulses. The theoretical treatment of Breizman {\it et al.} \cite{Breizman} predicts that, for very short pulses, the ion energy will scale as $\tau^2$, where $\tau$ denotes the laser pulse duration. It is clear that this scaling is not observed in the present experiments.

\acknowledgements
We are grateful to the Department of Science and Technology, Government of India for partial but important financial support for our femtosecond laser system.

\end{document}